\documentclass[graybox]{svmult}

\usepackage{amsmath, amssymb}
\usepackage{type1cm}        
\usepackage{makeidx}         
\usepackage{graphicx}        
\usepackage{multicol}        
\usepackage[bottom]{footmisc}

\usepackage{newtxtext}       %
\usepackage{newtxmath}       

\makeindex             

\newcommand{\R}{\mathbb R}

\def\be#1\ee{\begin{equation}#1\end{equation}}
\newcommand{\fer}[1]{(\ref{#1})}

\newcommand{\bq}{\begin{equation}}
\newcommand{\eq}{\end{equation}}

\def\bqa{\begin{eqnarray}}
\def\eqa{\end{eqnarray}}

\def\e{\epsilon}

\newcommand{\bd}{\begin{displaymath}}
\newcommand{\ed}{\end{displaymath}}
\newcommand{\ba}{\begin{eqnarray}}
\newcommand{\ea}{\end{eqnarray}}

\def\R{\mathbb{R}}

\newenvironment{equations}{\equation\aligned}{\endaligned\endequation}

\begin{document}

\title*{Statistical description of human addiction phenomena.}
\author{Giuseppe Toscani}
\institute{Giuseppe Toscani\at Department of Mathematics and IMATI Institute of CNR, Via Ferrata 5, 27100 Pavia, Italy. \email{giuseppe.toscani@unipv.it}}
%
%
\maketitle

\abstract*{ We study the evolution in time of the statistical distribution of some addiction phenomena in a system of individuals. The kinetic approach leads to build up a novel class of Fokker--Planck equations describing relaxation of the probability density solution towards a generalized Gamma density. A qualitative analysis reveals that the relaxation process is very stable, and does not depend on the parameters that measure the main microscopic features of the addiction phenomenon.  }

\abstract{ We study the evolution in time of the statistical distribution of some addiction phenomena in a system of individuals. The kinetic approach leads to build up a novel class of Fokker--Planck equations describing relaxation of the probability density solution towards a generalized Gamma density. A qualitative analysis reveals that the relaxation process is very stable, and does not depend on the parameters that measure the main microscopic features of the addiction phenomenon.}

\section{Introduction}
\label{sec:1}
A powerful way to describe the collective behavior of a multi-agent system of individuals with respect to a selected social phenomenon, whose intensity can be characterized by the parameter $x$, is to resort to kinetic theory, and to study the evolution of the number density $f=f(x,t)$ of individuals which are characterized by  the value $x\in \R_+$ at time $t\ge 0$, in terms of its changes in time by microscopic interactions (Boltzmann-type description \cite{NPT,PT13}). These interactions are built taking into account the main features of the phenomenon under study, which are quantified resorting to a precise mathematical description. 

This approach  has been originally applied to the statistical description of wealth distribution  in a multi-agent systems of individuals to better understand the reasons behind the formation of Pareto curves \cite{ChaCha00,CCM,ChChSt05,CPP,CoPaTo05,DY00,DMTb,TBD}. More recently, this modeling activity moved to social sciences, where, together with the investigation of opinion formation \cite{BN2,BN3,BN1,BeDe,Bou,Bou1,Bou2,CDT,DMPW,GGS,GM,Gal,GZ,SW,To1}, other aspects of modern societies, like conflicts, criminality, and city size formation have been considered \cite{BHT,BKS,BCKS, GT-ec,GT-city}. 

Since social phenomena are deeply based on behavioral aspects of agents, the microscopic kinetic interactions have been often modeled to reproduce these features.  To our knowledge, the first kinetic model in which   psychological and behavioral components of the agents have been explicitly considered has been proposed in \cite{MD} to model the price formation of a good in a multi-agent market, consisting of  two different trader populations. The kinetic description in \cite{MD} was inspired by the microscopic Lux--Marchesi model \cite{LMb,LMa} (cf. also \cite{LLS,LLSb}).  The microscopic trading rules of agents were assumed to depend both on the opinion of traders like in \cite{To1},  and on the way they interact with each other and perceive risks.  This last aspect has been done by resorting, in agreement with the pioneering prospect theory by Kahneman and Twersky \cite{KT,KT1},  to interactions involving a suitable \emph{value function}. 

Analogous microscopic mechanism has been considered in \cite{GT18}, where the choice of a particularly adapted value function  justified  the statistical shape of the service time distribution in a call center, and, in more generality, the formation of a number of social phenomena which can be described by a lognormal distribution \cite{GT19}. 

Starting from \cite{CoPaTo05}, where the formation of Pareto tails in the wealth distribution in a western society was studied at different scales, the analytical description of the stationary distribution of the social phenomenon under investigation was obtained resorting to a particular asymptotic limit of the kinetic equation of Boltzmann type, which results in a Fokker--Planck type equation  \cite{PT13,Vi}, still reminiscent of the microscopic interaction mechanism. This is particularly evident in the kinetic description of \cite{GT18,GT19}, where the shape of the value function to insert in the microscopic interaction is deeply connected to the steady state of the Fokker--Planck asymptotic equation.

The leading idea in \cite{GT18}  was recently applied to the study of the phenomenon of alcohol consumption in \cite{DT}. There, the choice of a new class of value functions, suitable to  model  the possibility of addiction in the microscopic interaction, led to a class of Fokker--Planck equations with a steady state given by a generalized Gamma distribution \cite{Lie,Sta}. 

The findings of \cite{DT} are in agreement with  the  exhaustive fitting analysis presented in \cite{Keh, Reh}. In these papers the analysis of the fitting of real data about alcohol consumption in a huge number of countries, pushed the authors to conclude that, among various probability distributions often used in this context, Gamma and Weibull distributions (particular cases of the generalized Gamma \cite{Sta}), appeared to furnish a better fitting with respect to the Log-normal distribution, first proposed by Ledermann \cite{Led} as a reasonable model for the consumption problem.

Let   $f = f(x,t)$  denote the probability density of individuals which are characterized by  an alcohol consumption value  equal to $x\in \R_+$ at time $t\ge 0$.  Then, the time evolution of the density $f$ is shown in \cite{DT} to obey to a linear Fokker--Planck equation, that  reads
\begin{equation}\label{FP}
 \frac{\partial f(x,t)}{\partial t} =  \frac{\partial^2 }{\partial x^2}
 \left(x^{2} f(x,t)\right )+ 
 \frac{\partial}{\partial x}\left[ \left(  \frac\delta{\theta^\delta} x^\delta \, -(\kappa + 1 )\right) x f(x,t)\right].
 \end{equation}
In equation \fer{FP} $ \theta, \kappa$ and $\delta$ are positive constants  related to the relevant characteristics of the phenomenon under study.  
Moreover, in all cases considered in \cite{DT}, in agreement with the fitting analysis in  \cite{Keh, Reh},  the constant $\delta \le 1$.  
The equilibrium state of the Fokker--Planck equation is given  the  generalized Gamma  density \cite{Lie,Sta}
 \be\label{equili}
f_\infty(x;\theta, \kappa,\delta) = \frac \delta{\theta^\kappa} \frac 1{\Gamma\left(\kappa/\delta \right)} x^{\kappa -1}
\exp\left\{ - \left( x/\theta\right)^\delta\right\}.
 \ee 
In this paper, we aim in improving the analysis of \cite{DT}, by generalizing it to different addiction phenomena, and by taking into account a new modeling assumption, in the spirit of the recent paper \cite{FPTT19}.  This leads to  describe the collective behavior of a multi-agent system of individuals subject to some addiction phenomena, whose intensity can be measured in terms of a positive parameter $x$, in terms of a new class of Fokker--Planck equations.
 
Given the probability density  $f = f(x,t)$  of individuals which are characterized by  an addiction value  equal to $x\in \R_+$ at time $t\ge 0$,  the time evolution of the density $f$ is shown to obey to a Fokker--Planck equation similar to \fer{FP}, that  now reads 
\begin{equation}\label{FPori}
 \frac{\partial f(x,t)}{\partial t} =  \frac{\partial^2 }{\partial x^2}
 \left(x^{2-\delta} f(x,t)\right )+ 
 \frac{\partial}{\partial x}\left[ \left(  \frac\delta{\theta^\delta} x\, -(\kappa +1-\delta)x^{1-\delta}\right) f(x,t)\right].
 \end{equation}
Similarly to equation \fer{FP}, $\theta$, $\kappa$ and $\delta$ are positive constants  related to the relevant characteristics of the phenomenon under study, and $\delta \le 1$.  The main difference between the Fokker--Planck equation \fer{FP} and the present one, is that both the coefficient of diffusion and the drift term are scaled by a factor $x^\delta$. This scaling has no effect on the steady state, so that the new equation \fer{FPori} has the same steady state \fer{equili} of equation \fer{FP}.

However, this new Fokker--Planck equation seems to be better adapted to describe the addiction phenomena, since it allows to obtain an explicit rate of relaxation of the solution towards the equilibrium.   Given an initial probability density $f_0(x)$ with a bounded variance, it can  be proven by classical entropy methods that the (unique) solution to the initial-boundary value problem for the Fokker--Planck equation \fer{FPori} converges towards the equilibrium density  \fer{equili} exponentially fast in time with explicit rate \cite{To2}, a result that seems not available for the solution to \fer{FP}.

The analysis of the present paper underlines the importance of the generalized Gamma density in the statistical description of social phenomena. Previous findings in this direction were concerned with event history and survival analysis \cite{box}.

Addiction phenomena which can be described by the Fokker--Planck equation \fer{FPori} include alcohol consumption \cite{DT}, on line gambling \cite{TTZ}, as well as the abuse of the insights of social networking sites \cite{KG}.  This new form of addiction is very recent, since online social networking sites reached  a very high popularity only in the last decade, involving more and more  individuals of the society to connect with others who share similar interests. The perceived need to be online was noticed to often result in compulsive use of these sites, which in extreme cases may produce symptoms and consequences traditionally associated with substance-related addictions \cite{KG}. 

In more details, in Section \ref{model} we will briefly describe the modeling assumptions of \cite{DT} and \cite{TTZ}, relative to the addiction phenomena of alcohol consumption and, respectively, to web gambling activity. In particular, we will outline the importance to resort to a variable collision kernel in the underlying linear Boltzmann equation.  A \emph{grazing collision} limit \cite{Vi} procedure finally allows to recover the Fokker--Planck equation \fer{FPori}. This will be the argument of Section \ref{fokker}.

A short review of the qualitative analysis of the Fokker--Planck equation \fer{FPori}, recently obtained in \cite{To2}, will be done in Section \ref{quali}.

\section{Kinetic description of addiction phenomena}
\label{model}

The goal of kinetic modeling  is to describe the collective behavior of a multi-agent system of individuals with respect to a certain hallmark by resorting to the typical elementary (microscopic) variations of the hallmark itself. In the case under investigation, the hallmark to be studied is the degree of addiction of the population of individuals relative for example to gambling, alcohol consumption or abuse of insights of social networking sites, measured by a variable x which varies continuously from $0$ to $+\infty$. 
 
Following the well-consolidated approach furnished by the kinetic theory \cite{FPTT, NPT, PT13}, the statistical description of the addiction variable will be described by resorting to a linear Boltzmann-type equation in which the unknown is the probability distribution $f= f(x,t)$ of the agents with a degree of addiction equal to $x$ at time $t \ge 0$. The kinetic model is built up by taking into account some basic hypotheses we enumerate below \cite{DT,TTZ}. 

To fix ideas, and to fully understand the main steps of the kinetic construction, we will refer to the description of the possible abuse of the insights of social networking sites in a society of individuals. In this case, we assume that the meaning of the variable $x$ is the weekly time (in seconds) spent to visit web sites. 
A key assumption is to consider the population homogeneous with respect to the phenomenon, assumption that requires to restrict it with respect to some characteristics, like age, sex and social class \cite{KG}. 

Once the homogeneity assumption is satisfied, individuals in the system can be considered indistinguishable \cite{PT13}, so that the state of any individual at any instant of time $t\ge 0$ is completely characterized by the time $x \ge0$ spent in web activities.  
The unknown is the density (or distribution function) $f = f(x, t)$, where $x\in \R_+$ and the time $t\ge 0$, and the target is to study its time evolution towards a certain equilibrium.

In general, the density function is normalized to one
\[
\int_{\R_+} f(x, t)\, dx = 1.
\]
The density changes in time since individuals connect (and disconnect) many times in the given period of a week, thus continuously upgrading the time $x$ spent in web activities. In agreement with the classical kinetic theory of rarefied gases, we will always denote a single upgrade of the quantity $x$ as a microscopic \emph{interaction}. 

In the phenomena under study, we will focus on two aspects, which appear to be common and essential in the eventual formation of addiction.

\begin{itemize}
\item{{\bf \emph{Assumption A}}:\,\,}
There is an entry level (represented by values of the variable $x$ below a certain value $\bar x$) that is accepted by most societies. The assumption of a moderate quantity of alcohol, an occasional gambling activity or a limited use of the mobile phone are indeed seen as completely normal. 
\item{{\bf \emph{Assumption B}}:\,\,}
There is an objective pleasure in spending time in these activities. Consequently, it is normally easier to increase the value of the quantity x than to decrease it. To prevent addiction, it is usual to fix an alarm level (represented by a suitable value $\bar x_L$ of the variable $x$, with $\bar x_L > \bar x$), that individuals should not exceed, and to continuously advertise about the dangers associated with addiction  values $x >\bar x_L$.
\end{itemize}
\emph{Assumption B}, strongly related to human behavior, has been fully considered in the kinetic modeling, at the level of  individual microscopic interactions, in various papers \cite{DT,GT18,GT19}, taking inspiration from the pioneering prospect theory by Kahneman and Twersky \cite{KT,KT1} and their representation of value functions. 

On the contrary, in \cite{DT,GT18,GT19} \emph{Assumption A}, mostly related to the collective behavior of the system of individuals, was not taken into account.
The mathematical translation of the entry level corresponds to assign a different value (frequency) to the elementary interactions in terms of the value $x$. A reasonable hypothesis is to assume that the frequency of interactions relative to a value $x$ of the addiction variable is inversely proportional to $x$. This relationship takes into account  both the highly probable access of individuals to the entry level, and the  rare possibility  to reach very high values of the $x$ variable. 

The choice of a variable interaction frequency  has been fruitfully applied in a different context \cite{FPTT19},  to better describe the evolution in time of the wealth distribution in a western society. There, the frequency of the economic transactions has been proportionally related to the values of the wealth involved, to take into account the low interest of trading agents in transactions with small values of the traded wealth.

As discussed in \cite{FPTT19}, the introduction of a variable kernel into the kinetic equation does not modify the shape of the equilibrium density, but it allows a better physical description of the phenomenon under study, including an exponential rate of relaxation to equilibrium for the underlying Fokker--Planck type equation. 
 
Following \cite{DT,GT18,GT19}, we will now illustrate the mathematical formulation of \emph{Assumption B}. The microscopic variation of time spent on social networks by individuals will be taken in the form
 \be\label{coll}
 x_* = x  - \Phi(x/\bar x_L) x + \eta x.
 \ee
In a single interaction the value $x$ of time can be modified for two reasons, expressed by two terms, both proportional to the value $x$. In the first one  the coefficient $\Phi(\cdot)$, which can assume assume both positive and negative values,  characterizes the predictable behavior of agents. The second term takes into account a certain amount of human unpredictability. The usual choice is to assume that the random variable $\eta$ is of zero mean and bounded variance, expressed by $\langle \eta \rangle =0$, $\langle \eta^2 \rangle  = \lambda$, with $\lambda >0$.
Small random variations of the interaction \fer{coll} will be expressed simply by multiplying $\eta$ by a small positive constant $\sqrt\e$, with $\e \ll 1$, which produces the new (small) variance $\e\lambda$.

The function $\Phi$ plays the role of the \emph{value function} in the prospect theory of Kahneman and Twersky \cite{KT, KT1}, and contains the mathematical details of the expected human behavior in the phenomenon under consideration, namely the fact that it is normally easier to increase the value of $x$ than to decrease it, in relationship with the alarm value $\bar x_L$.
In terms of the variable $ s = x/\bar x_L$ the value functions considered in \cite{DT} to describe alcohol consumption are given by
 \be\label{vd}
 \Phi_\delta^\e(s) = \mu \frac{e^{\e(s^\delta -1)/\delta}-1}{e^{\e(s^\delta -1)/\delta}+1 } , \quad  s \ge 0,
 \ee
where $0 < \delta \le 1$ and $0 <\mu <1$ are suitable constants characterizing the intensity of the individual behavior, while the constant $\e >0$ is related to the intensity of the interaction. Hence, the choice $\e\ll 1$ corresponds to small variations of the mean difference $\langle x_* -x\rangle$.   In \fer{vf}, the value $\mu$ denotes the maximal amount of variation of $x$ that agents will be able to obtain in a single interaction. Note indeed that the value function $\Phi_\delta^\e(s)$ is such that 
 \be\label{bounds}
  -\mu \le \Phi_\delta^\e(s) \le \mu.
 \ee
Clearly, the choice $\mu <1$ implies that, in absence of randomness, the value of $x^*$ remains positive if $x$ is positive. As proven in \cite{DT}, the value function satisfies 
 \be\label{ccd}
 -\Phi_\delta^\e\left(1-s \right) > \Phi_\delta^\e\left(1+s \right),
 \ee
 and 
 \be\label{cce}
 \frac d{ds}\Phi_\delta^\e\left(1+s \right)< \frac d{ds}\Phi_\delta^\e\left(1-s \right). 
 \ee
These properties are in agreement with the expected behavior of agents, since deviations from the reference point ($s=1$ in our case), are bigger below it than above. Letting $\delta \to 0$ in \fer{vd} allows to recover the value function
\be\label{vf}
 \Phi_0^\e(s) = \mu \frac{s^\e -1}{s^\e +1} , \quad  s \ge 0.
 \ee
introduced in \cite{GT18,GT19} to describe phenomena characterized by the lognormal distribution \cite{Aic,Lim}.  Hence, this choice is in full agreement with the data fitting of alcohol consumption proposed by Ledermann in 1956 \cite{Led}, choice which is still used in present times (cf. also the recent paper \cite{Mie} and the references therein).

Given the \emph{interaction}  \fer{coll}, for any choice of the value function $\Phi$ the study of the time-evolution of the distribution of the lenght $x$ of periods spent on web follows by
resorting to kinetic collision-like models \cite{Cer,PT13}. The variation of the  density $f(x,t)$  obeys to a linear
Boltzmann-like equation, fruitfully written
in weak form. The weak form corresponds to say that the solution $f(x,t)$
satisfies, for all smooth functions $\varphi(x)$ (the observable quantities)
 \begin{equation}
  \label{kin-w}
 \frac{d}{dt}\int_{\R_+}\varphi(x)\,f(x,t)\,dx  = 
  \Big \langle \int_{\R_+}\chi(x) \bigl( \varphi(x_*)-\varphi(x) \bigr) f(x,t)
\,dx \Big \rangle.
 \end{equation}
 Here expectation $\langle \cdot \rangle$ takes into account the presence of the random parameter $\eta$ in the microscopic interaction \fer{coll}. The function $\chi(x)$ measures the interaction frequency.

The right-hand side of equation \fer{kin-w} measures the variation in density between individuals that modify their value from $x$ to $x_* $ (loss term with negative sign) and   agents  that  change their value from  $x_*$ to  $x$  (gain term with positive sign).

In \cite{DT}, the simplification of the Maxwell molecules, leading to a constant interaction kernel  $\chi$, has been assumed.  This simplification, maybe not so well justified from a modeling point of view, is the common assumption in the Boltzmann-type description of socio-economic phenomena \cite{FPTT,PT13}. 

In \cite{FPTT19}, the Maxwellian assumption has been analyzed in its critical aspects. 
There, starting from a careful analysis of the microscopic economic transactions of the kinetic model, allowed to conclude that the choice of a constant collision kernel included as possible also interactions which human agents would exclude \emph{a priori}. This was evident for example in the case of interactions in which an agent that trades with a certain amount of wealth, does not receive (excluding the risk) some wealth back from the market. 

Following this line of thought, we can express the mathematical form of the kernel $\chi(x)$
by taking into account \emph{Assumption A}, which implies that the frequency of changes which leads to increase the amount of time $x$ is inversely proportional to $x$. Hence, in the addiction setting, it seems natural to consider collision kernels in the form
 \be\label{Ker}
 \chi(x) = \alpha x^{-\beta},
  \ee
for some constants $\alpha >0 $ and $\beta >0$.  This kernel assigns a low probability to happen to interactions in which individuals are subject to a high degree of addiction, and assigns a high probability to happen to interactions in which the value of the addiction variable $x$ is close to zero. 

The values of the constants $\alpha$ and $\beta$ can be suitably chosen by resorting to the following argument. For small values of the $x$ variable, the rate of growth of the value function \fer{vf} is given by
 \be\label{gro}
\frac d{dx} \Phi_\delta^\e\left(\frac x{\bar x_L}\right) \approx \mu \e \left(\frac x{\bar x_L}\right)^{\delta -1}. 
 \ee
This shows that, for small values of $x$, the mean individual growth predicted by the value function is proportional to $x^{\delta -1}$. Then, the choice $\beta = \delta$ would correspond to a collective growth independent of the parameter $\delta$ characterizing the value function.

A second important fact is that the individual rate of growth \fer{gro} depends linearly on the positive constant $\e$, and it is such that the intensity of the variation decreases as $\e$ decreases. Then, the choice
 \[
 \alpha = \frac\nu\e
 \]
is such that the collective growth remains bounded even in presence of very small values of the constant $\e$. 
With these assumptions, the weak form of the Boltzmann-type equation \fer{kin-w}, suitable to describe addiction phenomena, is given by
 \begin{equation}
  \label{kin}
 \frac{d}{dt}\int_{\R_+}\varphi(x)\,f_\e(x,t)\,dx  = \frac\nu\e\,\,
  \Big \langle \int_{\R_+}x^{-\delta} \bigl( \varphi(x_*)-\varphi(x) \bigr) f_\e(x,t)
\,dx \Big \rangle.
 \end{equation}
 Note that, in consequence of the choice made on the interaction kernel $\chi$, the evolution of the density $f_\e(x,t)$ is tuned by the parameter $\e$, which characterizes both the intensity of interactions and the interaction frequency.

\section{Fokker--Planck description and equilibria}
\label{fokker} 

For any choice of the value function \fer{vf}, the linear kinetic equation \fer{kin} describes the evolution of the density consequent to interactions of type \fer{coll}. The parameter $\e$ is closely related to the intensity of interactions. In particular, values $\e \ll 1$ describe the situation in which a single interaction  determines only an extremely small change of the value $x$. This situation is well-known in kinetic theory of rarefied gases, where interactions of this type are called \emph{grazing} collisions \cite{PT13,Vi}.  At the same time \cite{FPTT}, the balance of this smallness with the random part is achieved by setting 
 \be\label{scal}
  \eta \to \sqrt\e \eta.
 \ee
In this way the scaling assumptions allow to retain the effect of all parameters appearing in \fer{coll} in the  limit procedure. An exhaustive discussion on these scaling assumptions can be found in \cite{FPTT} (cf. also \cite{GT18} for analogous computations in the case of the Log-normal case). For these reasons, we address the interested reader  to these review papers for details.

Letting $\e \to 0$, the weak solution $f_\e(x,t)$ to  the kinetic model \fer{kin} converges towards $f(x,t)$, solution of a Fokker--Planck type equation \cite{FPTT}. 
Indeed, the limit density $f(x,t)$ is such that the time variation of the (smooth) observable quantity $\varphi(x)$  satisfies
 \begin{equations}\label{FP-weak}
 & \frac{d}{dt}\int_{\R_+}\varphi(x) \,f(x,t)\,dx  = \\
 & \nu \int_{\R_+} \left( - \varphi'(x)\,x^{1-\delta} \,\frac\mu{2\delta} \left( \left(\frac x{\bar x_L}\right)^\delta - 1 \right)   + \frac \lambda 2 \varphi''(x) x^{2-\delta} \right) f(x,t)\, dx 
 \end{equations}
Hence, provided the boundary terms produced by the integration by parts vanish,  equation \fer{FP-weak} coincides with the weak form of the Fokker--Planck equation
 \begin{equation}\label{FP2}
 \frac{\partial f(x,t)}{\partial t} = \nu \left[\frac \lambda 2 \frac{\partial^2 }{\partial x^2}
 \left(x^{2-\delta} f(x,t)\right )+ \frac \mu{2}
 \frac{\partial}{\partial x}\left( \frac 1\delta\left( \left(\frac x{\bar x_L}\right)^\delta - 1 \right) x^{1-\delta} \, f(x,t)\right)\right].
 \end{equation}
Equation \fer{FP2} describes the evolution of the distribution density $f(x,t)$ of the weekly time $x \in \R_+$ spent on social networks related activities in the limit of the \emph{grazing} interactions.  The steady state density can be explicitly evaluated \cite{DT}, and it results to be a generalized Gamma density, with parameters linked to the details of the microscopic interaction \fer{coll}.

Let us set $\gamma = \mu/\lambda$, and suppose that 
 \be\label{cc}
\gamma >\delta(1-\delta).
 \ee
Note that condition \fer{cc} is satisfied, independently of $\delta$, when $\mu \ge 4\lambda$, namely when the variance of the random variation in \fer{coll} is small with respect to the maximal variation of the value function.  Note that the smallness assumption \fer{cc} is typical of addiction phenomena, where individuals live their addiction without large unpredictable variations. Under condition \fer{cc}, it can be easily verified that  the  steady state solutions to \fer{FP2}  are given by the functions \cite{DT} 
 \be\label{equilibrio}
f_\infty(x) =  f_\infty(\bar x_L){\bar x}_L^{2-\delta} x^{\gamma/\delta +\delta -2}  \exp\left\{ - \frac \gamma{\delta^2}\left( \left( \frac x{\bar x_L} \right)^\delta -1 \right)\right\}.
 \ee 
By fixing the mass of the steady state \fer{equilibrio} equal to one,  the consequent probability density is the generalized Gamma $f_\infty(x;\theta, \kappa,\delta)$ defined by \fer{equili}, characterized in terms of the shape $\kappa>0$, the scale parameter $\theta >0$, and the exponent $\delta >0$ that in the present situation are given by 
 \be\label{para}
 \kappa =  \frac\gamma\delta +\delta -1,  \quad  \theta = \bar x_L \left( \frac{\delta^2}\gamma\right)^{1/\delta}.
 \ee
Clearly, condition \fer{cc} implies a positive value for the shape. The limit $\delta \to 0$ in the Fokker--Planck equation \fer{FP2} corresponds to the drift term induced by the value function \fer{vf}. In this case, the equilibrium distribution \fer{equilibrio} takes the form of a lognormal density \cite{GT18}. 
 
Note that for all values $\delta >0$ the moments are expressed in terms of the parameters  denoting respectively the alarm level $\bar x_L$, the variance $\lambda$ of the random effects and the values $ \delta$ and $\mu$ characterizing  the value function $\phi_\delta^\e$ defined in \fer{vd}. 

Going back to the fitting analysis presented in \cite{Keh, Reh}, who led to identify as correct statistical distributions for alcohol consumption the Gamma and Weibull ones, we recall that these cases are obtained by choosing $\delta =1$ and $\delta = \kappa$, respectively. In particular, the case of Gamma distribution leads to a mean value of the addiction  equal to the alarm level $\bar x_L$, while for the Weibull distribution, where $\gamma =\delta$, the mean value of the addiction is given by
\be\label{mean-weibull}
\int_{\R_+} x\, f_\infty(x) \,dx = \bar x_L \delta^{1/\delta -1}\Gamma\left(   \frac 1\delta \right).
\ee
Note that, since $\delta <1$, in this case the mean value is strictly less than the alarm level $\bar x_L$. If for example $\delta =1/2$, the mean value is equal to $\bar x_L/2$. Hence, the Weibull case corresponds to the situation in which the addiction phenomenon is sensible to the advertisements about possible dangers. In the general case, the mean takes the value
\be\label{mean}
\int_{\R_+} x\, f_\infty(x) \,dx = \bar x_L \left(\frac{\delta^2}\gamma\right)^{1/\delta}\frac{\Gamma\left[\frac 1\delta\left(\frac\gamma\delta +\delta\right)\right]}{\Gamma\left[\frac 1\delta\left(\frac\gamma\delta +\delta -1\right)\right]}.
\ee
Exact computation of the mean can be done by choosing, for $1/2 <\delta <1$, the value  $\gamma = \delta^2$. This choice is such that condition \fer{cc} is satisfied. In this case 
 \[
\int_{\R_+} x\, f_\infty(x) \,dx = \bar x_L \frac 1{\Gamma\left(2 -\frac 1\delta \right)} > \bar x_L. 
 \]
This shows that the alarm level can be exceeded in the presence of a small variance of the random variation (with respect to the maximal variation of the value function), which corresponds to a strong addiction phenomenon.

\section{Relaxation to equilibrium}
\label{quali}

Scaling time in the Fokker--Planck equation \fer{FP2} allows to write it in the clean form \fer{FPori}, which outlines the dependence  of both the diffusion and drift terms on the shape $\kappa>0$, the scale parameter $\theta >0$, and the exponent $\delta >0$. Also, equation \fer{FPori} allows to directly recover the generalized Gamma density \fer{equili} in terms of the same parameters.  
It has to be remarked once more that  the limit procedure described in Section \ref{fokker}  leads to equation \fer{FP-weak}, namely to a weak version of the Fokker--Planck equation \fer{FP2}. Then, suitable boundary conditions have to be considered, to guarantee the equivalence of the two equations, and consequently the correct evolution of the main macroscopic quantities, and among them the mass conservation. The most used are the so-called \emph{no--flux} boundary conditions, expressed by 
\be\label{bc1}
\left. \frac{\partial }{\partial x}\left( x^{2-\delta} f(x,t)\right) + \left(  \frac\delta{\theta^\delta} x\, -(\kappa +1-\delta)x^{1-\delta}\right) \,f(x,t) \right|_{x=0,+\infty} = 0, \quad t>0.
\ee
In presence of the no-flux boundary conditions \fer{bc1} one can study, without loss of generality,  the initial-boundary value problem for equation \fer{FPori} with a probability density function, say $f_0(x)$, as initial datum.  Then, mass conservation implies that the solution $f(x,t)$ remains a probability density for all subsequent times $t >0$.

The qualitative analysis of the Fokker--Planck equations \fer{FPori} has been done in the recent paper \cite{To2}. There, the analysis was extended to values of $\delta$ in the interval $0< \delta \le 2$, thus covering generalized Gamma densities that range from the Log-normal density, corresponding to $\delta \to 0$, to Chi-densities, obtained for $\delta =2$. 

In the following, for the sake of completeness, we will briefly present the main results obtained in \cite{To2}, as well as the main properties of this class of Fokker--Planck equations.  As extensively discussed in \cite{FPTT}, Fokker--Planck type equations of type \fer{FPori} can be rewritten in different equivalent forms, each one useful for various purposes. 
For given $t >0$, let
 \be\label{dist}
 F(x,t) = \int_0^x f(y,t)\, dy
 \ee
denote the probability distribution induced by the probability density $f(x,t)$, solution of the Fokker--Planck equation \fer{FPori}.
In \cite{To2}, the writing  the Fokker--Planck equations \fer{FPori} in terms of the distribution $F(x,t)$, highlighted an interesting feature of their solutions. 
 
Integrating both sides of equation \fer{FPori} on the interval $(0,x)$, and applying condition \fer{bc1} on the boundary $x=0$, it is immediate to verify that $F(x,t)$ satisfies the equation 
\begin{equation}\label{FP3}
 \frac{\partial F(x,t)}{\partial t} =  x^{2-\delta}\, \frac{\partial^2 }{\partial x^2}
  F(x,t) + \left(  \frac\delta{\theta^\delta} x\, -(\kappa -1)x^{1-\delta}\right) 
 \frac{\partial}{\partial x}F(x,t),
 \end{equation}
The no-flux boundary conditions \fer{bc1} then guarantee that, for any $t \ge 0$  
 \be\label{limo}
 F(0,t) = 0; \qquad \lim_{x \to +\infty} F(x,t) = 1.
 \ee
The second condition in \fer{limo}  corresponds to mass conservation. 
 
Given a positive constant $m >0$, let us consider the  transformation
\be\label{key}
 F(x,t) = G(y,\tau), \qquad y =y(x) = x^m, \quad \tau = \tau(t) = m^2 t.
\ee
Then it holds 
 \[
 \frac{\partial}{\partial t}F(x,t) = m^2 \,\frac{\partial}{\partial \tau}G(x,\tau),   
  \]
while
 \[
 \frac{\partial}{\partial x}F(x,t) =  m x^{m-1}\, \frac{\partial}{\partial y}G(y,\tau), 
 \]
and
\[
\frac{\partial^2}{\partial x^2}F(x,t) =  m^2x^{2m-2}\, \frac{\partial^2}{\partial y^2}G(y,\tau) + m(m-1) x^{m-2}\, \frac{\partial}{\partial y}G(y,\tau).
\]
Hence, substituting into \fer{FP3} the above identities and using the inverse relation $x = y^{1/m}$, it follows that $G(y, \tau)$ satisfies the Fokker--Planck equation
\begin{equation}\label{FP4}
 \frac{\partial G(y,\tau)}{\partial \tau} =  y^{2-\delta/m}\, \frac{\partial^2 }{\partial y^2}
  G(y,\tau) + \left(  \frac{\delta/m}{(\theta^m)^{\delta/m}} y\, -\left(\frac\kappa{m} -1\right)y ^{1-\delta/m}\right) 
 \frac{\partial}{\partial y} G(y,\tau).
 \end{equation}
 Moreover, if $F(x,t)$ satisfies conditions \fer{limo} for any $t \ge0$, $G(y,\tau)$ still satisfies the same conditions  for any $\tau \ge 0$.
  
Note that equation \fer{FP4} has the same structure of equation \fer{FP3}, with the constants $\kappa$,  $\theta $, and  $\delta $ substituted by $\theta^m$, $\kappa/m$  and  $\delta/m$ . Consequently, its equilibrium distribution is given by the generalized Gamma density 
 \be\label{equi2}
 f_\infty\left(y;\theta^m, \frac\kappa{m},\frac\delta{m}\right).
 \ee
It is interesting to remark that, if $X(t)$ is the random process with probability distribution given by $F(x,t)$, by construction $G(x,t)$ is the probability distribution of the process $X^m(t/m^2)$. 

In  \cite{Sta}, Stacy noticed that the generalized Gamma densities satisfy a similar property. Given a constant $m >0$, when a random variable is distributed according to \fer{equili},  $X^m$ is distributed according to \fer{equi2}. 

Using this property, in \cite{To2}  two special cases, corresponding to the choices $m =\delta$ and $m =\delta/2$, were considered. These cases correspond to simplify the drift term and the diffusion coefficient, respectively. Indeed, the choice $m =\delta$  leads to the Fokker--Planck equation with linear drift
\begin{equation}\label{FP-1}
 \frac{\partial f(x,t)}{\partial t} =  \frac{\partial^2 }{\partial x^2}
 \left(x f(x,t)\right )+ 
 \frac{\partial}{\partial x}\left[ \left(  \frac x{\theta^\delta} \, -\frac\kappa\delta \right) f(x,t)\right].
 \end{equation}
 The steady state of equation \fer{FP-1} is the standard Gamma distribution of  shape $\bar \kappa= \kappa/\delta$ and scale $\bar\theta= \theta^\delta$
 \be\label{gamma}
f_\infty(x;\theta^\delta, \kappa/\delta, 1) = f_\infty(x;\bar\theta, \bar\kappa, 1)\frac 1{{\bar\theta}^{\bar\kappa}} =  \frac 1{\Gamma\left(\bar\kappa \right)} x^{\bar\kappa -1}
\exp\left\{ -x/\bar\theta\right\}.
 \ee 
Likewise, the choice $m=\delta/2$ leads to the Fokker--Planck equation with constant coefficient of diffusion 
\begin{equation}\label{FP-2}
 \frac{\partial f(x,t)}{\partial t} =  \frac{\partial^2 }{\partial x^2} f(x,t)+ 
 \frac{\partial}{\partial x}\left[ \left(  \frac2{\theta^\delta} x\, -\left(\frac {2\kappa}\delta -1\right)x^{-1}\right) f(x,t)\right].
 \end{equation}
In this second case, the steady state of equation \fer{FP-2} is the Chi-distribution of  shape $\tilde\kappa= 2\kappa/\delta$ and scale $\tilde\theta= \theta^{\delta/2}$
 \be\label{chi}
f_\infty(x;\theta^\delta, \kappa/\delta, 2) = f_\infty(x;{\tilde\theta}^2, \tilde\kappa/2, 2)= \frac 2{{\tilde\theta}^{\tilde\kappa/2}} \frac 1{\Gamma\left(\tilde\kappa/2 \right)} x^{\tilde\kappa -1}
\exp\left\{ -x^2/{\tilde\theta}^2\right\}.
 \ee 
Equation \fer{FP-1} has been exhaustively studied in a pioneering paper by Feller  \cite{Fel1},  who studied the initial boundary value problem with no-flux boundary conditions \fer{bc1}, and proved existence and uniqueness of solutions, positivity and mass conservation. 

It is interesting to remark that, when the shape $\bar\kappa=\kappa/\delta >1$,  there exists a positive and norm preserving solution of the initial-boundary value problem such that both it and its flux vanish at $x = 0$ \cite{Fel1}. This means that  when  $\kappa/\delta >1$ the boundary $x = 0$ acts both as absorbing and reflecting barrier and that no homogeneous boundary conditions need to be imposed. Mass conservation holds even without no flux boundary conditions.

In view of the aforementioned connections among the Fokker--Planck equations \fer{FPori}, the existence and uniqueness results relative to the exponent $\delta =1$ still hold for the initial-boundary value problem for equation \fer{FPori} characterized by a parameter $\delta \not=1$. For a given initial probability density $f_0(x)$, there exists a unique positive and mass preserving solution in presence of boundary conditions \fer{bc1}. Moreover,  if $\kappa > \delta$,  there exists a unique positive and norm preserving solution of the initial value problem such that both it and its flux vanish at $x = 0$.  Mass conservation holds even without no flux boundary conditions.

The Fokker--Planck equation \fer{FP-2}, with constant diffusion coefficient, allows to prove, using the strategy of Otto and Villani \cite{OV}, that, provided $\kappa \ge \delta/2$, the generalized Gamma densities \fer{equili} satisfy the weighted logarithmic Sobolev inequality
 \be\label{LS-G}
  H(f|f_\infty(\theta, \kappa,\delta)) \le \frac{\theta^\delta}{\delta^2} I_{2-\delta}(f|f_\infty(\theta, \kappa,\delta)),
 \ee
where, given two probability densities $f(x)$ and $g(x)$, with $x \in \R_+$, $H(f,g)$ denotes  the Shannon entropy of $f$ relative to  $g$
 \[
H(f|g) = \int_{\R_+} f(x) \log \frac{f(x)}{g(x)}\, dx , 
 \]
and, for a given constant $\beta \ge0$,  $I_\beta(f,g)$ denotes the weighted Fisher information of $f$ relative to $g$
 \[
 I_\beta(f|g) = \int_{\R_+} x^\beta f(x) \left(\frac d{dx}\log \frac{f(x)}{g(x)}\right)^2\, dx.
 \]
Inequality \fer{LS-G} then implies exponential convergence in relative entropy of the solution to the Fokker--Planck equation \fer{FPori} at the explicit rate $\theta^\delta/\delta^2$. Note that, provided  $\kappa \ge \delta/2$, the convergence rate does not depend on the value of $\kappa$.

 When the parameter $\theta$ is given by \fer{para}, the constant in the weighted logarithmic Sobolev inequality \fer{LS-G} takes the value
 \[
 \frac{\theta^\delta}{\delta^2} = \frac{{\bar x_L}^\delta}\gamma = \frac{{\bar x_L}^\delta\, \lambda}\mu.
 \]
Note that the rate of exponential convergence towards the equilibrium density increases with the alarm level $\bar x_L$ and with the variance $\lambda$ of the stochastic part of the microscopic interaction, while it decreases with respect to the maximal amount of variation $\mu$ of the value function \fer{vd}. Also, the behaviour with respect to the parameter $\delta$ that characterizes the value function \fer{vd} is different depending of the value of $\bar x_L$. The rate of convergence decreases with $\delta$ if $\bar x_L<1$, while it increases in the opposite situation. It is remarkable that there is no dependence on $\delta$ when the alarm level $\bar x_L=1$.

 \section{Conclusions}
Recent results on fitting of the statistical distribution of addiction phenomena in a multi-agent system \cite{Keh, Reh} lead to conjecture that these phenomena are well represented by a generalized Gamma distribution. In this paper we show that this type of probability distributions can be obtained as steady states of Fokker--Planck equations modeling addiction phenomena in terms of suitable microscopic interactions. A qualitative analysis of these equations then verifies that equilibrium is reached exponentially in time, thus justifying the fitting analysis.

\begin{acknowledgement}
This work has been written within the activities of GNFM (Gruppo Nazionale per la Fisica Matematica) of INdAM (Istituto Nazionale di Alta Matematica), Italy.
The research was partially supported by the Italian Ministry of Education, University and Research (MIUR) through the ``Dipartimenti di Eccellenza'' Programme (2018-2022) -- Department of Mathematics ``F. Casorati'', University of Pavia  and through the MIUR  project PRIN 2017TEXA3H ``Gradient flows, Optimal Transport and Metric Measure Structures''.
\end{acknowledgement}


\begin{thebibliography}{99.}%


\bibitem{Aic}
Aitchison, J.; Brown, J.A.C.: \emph{The Log-normal Distribution},
Cambridge University Press, Cambridge, UK 1957.

\bibitem{BHT}
Bellomo, N.; Herrero, M.A.; Tosin, A.: On the dynamics of social conflicts looking
for the Black Swan, {Kinet. Relat. Models} (6) 459--479 (2013)

\bibitem{BKS}
Bellomo, N.; Knopoff.; Soler, J.:
\newblock On the difficult interplay between life, complexity, and
  mathematical sciences,
\newblock { Math. Models Methods Appl. Sci.}  {\bf 23}  1861--1913 (2013)

\bibitem{BCKS}
Bellomo, N.; Colasuonno, F.; Knopoff.; Soler, J.: From a systems theory of
sociology to modeling the onset and evolution of criminality, {Netw. Heterog. Media}
\textbf{10}  421--441 (2015)

\bibitem{BN2}
 Ben-Naim, E.;  Krapivski, P.L.;  Redner, S.: Bifurcations and patterns in compromise processes, {Physica D} \textbf{183}  190--204  (2003)

\bibitem{BN3}
Ben-Naim, E.;  Krapivski, P.L.;  Vazquez, R.; Redner, S.: Unity and discord in opinion dynamics, {Physica A} \textbf{330}
 99--106  (2003)

\bibitem{BN1}
Ben-Naim, E.: Opinion dynamics: rise and fall of political parties, {Europhys. Lett.} \textbf{69} 671--677 (2005) 

\bibitem{BeDe}
 Bertotti, M.L.; Delitala, M.: On a discrete generalized kinetic approach for mo\-del\-ling persuader's influence in opinion formation processes,  {Math. Comp. Model.} \textbf{48} 
 1107--1121 (2008)

\bibitem{Bou}
Boudin, L.; Salvarani, F.: The quasi-invariant limit for a kinetic model of sociological collective behavior, {Kinetic Rel. Mod.} \textbf{2}  433--449 (2009)

\bibitem{Bou1}
 Boudin, L.; Salvarani, F.: A kinetic approach to the study of opinion formation, {ESAIM: Math. Mod. Num. Anal.} \textbf{43} 
  507--522 (2009)

\bibitem{Bou2}
Boudin, L.; Mercier, A.; Salvarani, F.: Conciliatory and contradictory dynamics in opinion formation, {Physica A} \textbf{391}  5672--5684 (2012)

\bibitem{box}
Box-Steffensmeier, J.M.; Jones, B.S.: \emph{Event History Modeling
A Guide for Social Scientists}. Cambridge University Press, Cambridge UK 2004

\bibitem{Cer}
Cercignani, C.:
\emph{The Boltzmann equation and its applications},
\newblock  Springer Series in Applied Mathematical Sciences,
  Vol. \textbf{67} Springer--Verlag, New York 1988.
  
\bibitem{ChaCha00} 
Chakraborti, A.; Chakrabarti, B.K.: Statistical
  Mechanics of Money: Effects of Saving Propensity, {Eur. Phys. J. B}
  \textbf{17} 167--170  (2000)    


\bibitem{CCM} 
Chatterjee, A.; Chakrabarti, B.K.; Manna, S.S.:
   Pareto law in a kinetic model of market with random saving propensity,
  {Physica A } {\bf 335}  155--163 (2004)

\bibitem{ChChSt05} 
Chatterjee, A.; Chakrabarti, B.K.; Stinchcombe, R.B.: Master equation for a kinetic model of trading  market and its analytic solution, {Phys. Rev. E} \textbf{72}  026126   (2005)


  \bibitem{CDT}
Comincioli, V.; Della Croce, L.; Toscani, G.:  A Boltzmann-like equation for choice formation, {Kinetic Rel. Mod.} \textbf{2}   135--149 (2009)

\bibitem{CPP}
Cordier, S.; Pareschi, L.; Piatecki, C.: Mesoscopic modelling of financial markets, {J. Stat. Phys.} \textbf{134} (1)   161--184 (2009)

\bibitem{CoPaTo05} 
Cordier, S.; Pareschi, L.; Toscani, G.: On a kinetic
  model for a simple market economy, { J. Stat. Phys.} \textbf{120}   253--277 (2005)
  

\bibitem{DT}
Dimarco, G.; Toscani, G.: Kinetic modeling of alcohol consumption.  arXiv:1902.08198 (2019) 
    
\bibitem{DY00}
Dr\v{a}gulescu, A.; Yakovenko, V.M.: {Statistical mechanics of money}, { Eur. Phys. Jour. B} \textbf{17}   723--729 (2000)

\bibitem{DMPW}
D{\"u}ring, B.;  Markowich, P.A.;  Pietschmann, J-F.;  Wolfram, M-T.:
\newblock Boltzmann and {F}okker-{P}lanck equations modelling opinion formation
  in the presence of strong leaders,
\newblock { Proc. R. Soc. Lond. Ser. A Math. Phys. Eng. Sci.}
  {\bf 465}  3687--3708 (2009) 

\bibitem{DMTb} 
D\"uring, B.; Matthes, D.; Toscani, G.: Kinetic equations modelling wealth redistribution:
a comparison of approaches,  {Phys.\ Rev.\ E} \textbf{78}   056103  (2008)

\bibitem{Zua}
Escobedo, M.;  Zuazua, E.: Large time behavior for convection-diffusion equations in $\R^N$, J. Funct. Anal. \textbf{100}  119--161 (1991)

\bibitem{Fel1}
Feller,  W.: Two singular diffusion problems, \emph{Ann. Math.} \textbf{54} (2)  173--182  (1951)
  


\bibitem{FPTT}
Furioli, G.; Pulvirenti, A.; Terraneo, E.;  Toscani, G.:  Fokker--Planck equations in the modelling of socio-economic phenomena, {Math. Mod. Meth. Appl. Scie.} \textbf{27} (1) 115--158 (2017)

\bibitem{FPTT19}
Furioli, G.; Pulvirenti, A.; Terraneo, E.;  Toscani, G.: Non-Maxwellian kinetic equations modeling the evolution of wealth distribution. (preprint) (2019)

\bibitem{GGS}
Galam, S.;  Gefen, Y.; Shapir, Y.: Sociophysics: A new approach of sociological collective behavior. I. Mean-behaviour description of a strike, { J. Math. Sociology} \textbf{9}  1--13 (1982)

\bibitem{GM}
Galam, S.; Moscovici, S.: Towards a theory of collective phenomena: consensus and attitude changes in groups, {Euro. J. Social Psychology} \textbf{21}  49--74 (1991)

\bibitem{Gal}
Galam, S.: Rational group decision making: A random field Ising model at $T= 0$. {Physica A} \textbf{238} 66--80  (1997) 

\bibitem{GZ}
Galam, S.;   Zucker, J.D.: From individual choice to group decision-making. {Physica A}  \textbf{287}  644--659 (2000)

\bibitem{GT-ec}
Gualandi, S.; Toscani, G: Pareto tails in socio-economic phenomena: a kinetic description. {Economics} \textbf{12}   1--17 (2018-31)
 
\bibitem{GT18}
Gualandi, S.; Toscani, G: Call center service times are lognormal. A Fokker--Planck description.
{Math. Mod. Meth. Appl. Scie.} \textbf{28}  (08) 1513--1527  (2018) 

\bibitem{GT19} 
Gualandi, S.; Toscani, G: Human behavior and lognormal distribution. A kinetic description.  Math. Mod. Meth. Appl. Scie. \textbf{29} (4) 717--753 (2019) 

\bibitem{GT-city}
Gualandi, S.; Toscani, G: The size distribution of cities: A kinetic explanation. Physica A \textbf{524}, 221--234 (2019)

\bibitem{KT}
Kahneman, D.; Tversky, A.:Prospect theory: an analysis of decision under risk,
{Econometrica} \textbf{47} (2)  263--292 (1979)


\bibitem{KT1}
Kahneman, D.; Tversky, A.: \emph{Choices, values, and frames}, Cambridge University Press, Cambridge, UK 2000.

\bibitem{Keh}
Kehoe, T.; Gmel, Gerritt; Shield,  K.D.; Gmel Gerhard;  Rehm, J.: Determining the best population-level alcohol consumption model and its impact on estimates of alcohol-attributable harms. {Population Health Metrics} \textbf{10}:6 (2012)

\bibitem{KG}
Kuss, D.J.; Griffiths, M.D.: Online social networking and addiction-A review of the psychological literature. Int. J. Environ. Res. Public Health , \textbf{8},  3528--3552 (2011)

\bibitem{Led}
Ledermann, S.: \emph{Alcool, Alcoolisme, Alcoolisation}, Vol. I. Presses Universitaires de
France, Paris (1956).

\bibitem{LLS}
 Levy, M.; Levy, H.; Solomon, S.: A microscopic model of the stock market:
Cycles, booms and crashes, {Econ. Lett.}  \textbf{45} 103--111 (1994) 

\bibitem{LLSb}
 Levy, M.; Levy, H.; Solomon, S.: {\it Microscopic simulation of financial markets: from
investor behaviour to market phenomena}, Academic Press,  San Diego, CA 2000

\bibitem{Lie} 
Lienhard, J.H.;  Meyer, P.L.: A physical basis for the generalized Gamma distribution.
{Quarterly of Applied Mathematics}, \textbf{25} (3)  330--334 (1967)

\bibitem{Lim}
Limpert, E.; Stahel,  W.A.;  Abbt, M.:
Log-normal distributions across the sciences: keys and clues, {BioScience} \textbf{51} (5)  341--352 (2001)

\bibitem{LMb}
Lux T.; Marchesi, M.: Scaling and criticality in a stocastich multi-agent
model of a financial market, {Nature} \textbf{397} (11)  498--500 (1999)


\bibitem{LMa}
Lux T.; Marchesi, M.: Volatility clustering in financial markets: a
microscopic simulation of interacting agents, {International Journal
of Theoretical and Applied Finance} \textbf{3}  675--702 (2000)

\bibitem{MD}
Maldarella, D.; Pareschi, L.: Kinetic models for socio--economic dynamics
of speculative markets, {Physica A} \textbf{391}  715--730 (2012)


\bibitem{Mie}
Mielecka-Kubien, Z.: On the estimation of the distribution of alcohol
consumption, {Mathematical Population Studies}, \textbf{25} (1)  1--19  (2018)

\bibitem{NPT}
Naldi, G.; Pareschi, L.; Toscani, G. eds.: \emph{Mathematical modeling of
collective behavior in socio-economic and life sciences},  Birkhauser,
Boston 2010.

\bibitem{OV}
Otto, F.; Villani, C.: Generalization of an inequality by Talagrand and links with the logarithmic Sobolev inequality, J. Funct. Anal. \textbf{173} 361--400 (2000)
  
\bibitem{PT13}
Pareschi, L.; Toscani, G.: \emph{Interacting multiagent systems: kinetic equations and Monte Carlo methods}, Oxford University Press, Oxford 2014  


\bibitem{Reh}
Rehm, J.; Kehoe, T.; Gmel, Gerritt; Stinson, F.; Grant, B.; Gmel Gerhard:
Statistical modeling of volume of alcohol exposure for epidemiological studies of
population health: the US example. {Population Health Metrics} \textbf{8}:3 (2010)


\bibitem{SW}
Sznajd--Weron, K.; Sznajd, J.: Opinion evolution in closed community,  {Int. J. Mod. Phys. C} \textbf{11}
  1157--1165 (2000)

\bibitem{Sta}
 Stacy, E.W.: A generalization of the Gamma distribution. {Ann. Math. Statist.} \textbf{33}  1187--1192 (1962)
 
\bibitem{To1}
Toscani, G.: Kinetic models of opinion formation, \emph{Commun.
Math. Sci.} \textbf{4}   481--496 (2006)

\bibitem{To2}
Toscani, G.: Entropy-type inequalities for generalized Gamma densities. Ricerche di Matematica (in press) (2020)

\bibitem{TTZ} Toscani, G.; Tosin, A.; Zanella M.:, Multiple-interaction kinetic modelling of a virtual-item gambling economy.  Phys. Rev. E \textbf{100} 012308 (2019)

\bibitem{TBD} 
Toscani, G.; Brugna, C.; Demichelis, S.: Kinetic models for the trading of goods, {J. Stat. Phys}, \textbf{151}   549--566 (2013)


\bibitem{Vi}
 Villani, C.:
   Contribution {\`a} l'{\'e}tude math{\'e}matique des
  {\'e}quations de {B}oltzmann et de {L}andau en th{\'e}orie cin{\'e}tique des
  gaz et des plasmas.  {\em PhD thesis, Univ. Paris-Dauphine} (1998)



\end{thebibliography}
\end{document}